\documentclass[aps,prstab,groupedaddress,showpacs,twocolumn,showkeys]{revtex4-1}
\usepackage{graphicx}
\usepackage{url}
\usepackage{amsmath}
\usepackage{color}
\begin{document}

\title{Characterization and control of linear coupling using turn-by-turn
beam position monitor data in storage rings}

\author{Yongjun Li}
\email{yli@bnl.gov}
\author{Lingyun Yang} 
\author{Weixing Cheng}

\affiliation{Brookhaven National Laboratory, Upton, NY-11973}

\date{\today}

\begin{abstract}
We introduce a new application of measuring symplectic generators to
characterize and control the linear betatron coupling in storage
rings. From synchronized and consecutive BPM~(Beam Position Monitor)
turn-by-turn~(TbT) readings, symplectic Lie generators describing the
coupled linear dynamics are extracted. Four plane-crossing terms in the
generators directly characterize the coupling between the horizontal and
the vertical planes. Coupling control can be accomplished by utilizing the
dependency of these plane-crossing terms on skew quadrupoles. The method
has been successfully demonstrated to reduce the vertical effective
emittance down to the diffraction limit in the newly constructed National
Synchrotron Light Source II~(NSLS-II) storage ring. This method can be
automatized to realize linear coupling feedback control with negligible
disturbance on machine operation.
\end{abstract}


\maketitle

\section{\label{intro}Introduction}

Linear betatron coupling due to tilting normal quadrupoles, vertical
displacement of beam orbit through sextupoles, or existence of skew
quadrupoles for vertical dispersion control, can directly degrade machine
performance in circular accelerators, such as electron storage rings used
as high-brilliance x-ray light sources. Accurate characterization and
control of the linear betatron coupling is of primary importance to
improve the brilliance for a low-emittance light source. It also provides
an important tool for beam volume and intensity related studies.

Thanks to the modern beam diagnostics techniques, turn-by-turn (TbT) beam
position monitor (BPM) data are available for most of synchrotron
radiation light sources now. From TbT data, a quadratic Lie generator can
be extracted by fitting linear one-turn-maps at each BPM location. The
coupling between two transverse planes can be directly characterized by
four plane-crossing terms in the generators. By using a set of skew
quadrupoles at non-dispersive sections, the coupling can be well
controlled without blowing up vertical dispersion. This method has been
successfully applied to the newly constructed National Synchrotron Light
Source II~(NSLS-II) storage ring.

There already exist many different analyses of dealing with linear
coupling~\cite{Edwards:1973fm,Mais:1982hx,Willeke:1988zu,sagan_2000,
  Sagan_Rubin_1999,Lebedev:2010zzc,Guignard:1975yu,Fischer_2003,
  Cai_2003,Luo_2004,Aiba_2012,Franchi_2007,Franchi_2011,Masaki_2009,
  Calaga:2005bn}. Some of them are well-known in our community, such as,
Teng-Edwards parameterization (symplectic matrix
normalization~\cite{Edwards:1973fm}), Mais-Ripken parameterization
(generalized Courant-Snyder parameterization with four $\beta$
functions~\cite{Mais:1982hx,Willeke:1988zu,Lebedev:2010zzc}), and
Guignard's perturbation theory~\cite{Guignard:1975yu}. In practice,
coupling correction is also carried out by minimizing the off-diagonal
orbit response matrix~\cite{Aiba_2012}, which is also a fitting module in
LOCO~\cite{Safranek_1997}. The purpose of this paper is to emphasize an
algorithm using two neighboring BPMs turn-by-turn data to extract the
coupling term in their one-turn generators, then to realize coupling
control with skew quadrupoles. Benefited from previous analyses, we can
demonstrate that direct minimization of coupling terms in the generators
is equivalent to these methods. Since TbT data can be accessed within
seconds, then skew quadrupole correction scheme can be automatized as a
feedback with a negligible interruption on machine routine operation. For
ring-based light sources, when insertion devices gaps are being closed or
opened, the coupling caused by the imperfection of those devices can be
under control.

The paper is organized in the following way: Sect.~\ref{coupling}
introduces coupled 2-dimensional linear Hamiltonian
dynamics. Sect.~\ref{measurement} describes the detailed procedure of
extracting one-turn-maps, and then Lie generators from beam TbT
data. Sect.~\ref{algorithm} explains the algorithm of implementing
coupling control or correction. In Sect.~\ref{exper}, we demonstrate its
application on the NSLS-II storage ring. A brief summary is given in
Sect.~\ref{summary}.

\section{\label{coupling}Linear coupling}
Consider a 2-dimensional (4-dimensinal phase space) coupled linear
periodic dynamical system, such as a charged particle traveling in a
storage ring. Particle coordinates in the phase space are denoted by
$\vec{v}(s)=(x,p_x,y,p_y)^T$, the 4 dimensional vector with the positions
and momenta at location $s$. Here $\mathbf{x}^T$ means the transpose of
vector or matrix $\mathbf{x}$. The one-turn-map $\mathbf{R}$ transforms
the vector $\vec{v}^{(n)}$ at turn $n$ to $\vec{v}^{(n+1)}$ at turn $n+1$
\begin{equation}\label{z=Mz}
\vec{v}(s)^{(n+1)} = \mathbf{R}(s)\vec{v}(s)^{(n)}.
\end{equation}
$\mathbf{R}(s)$ is a $4\times{}4$ matrix observed at the location of $s$
and can be written as.
\begin{equation}\label{M}
\mathbf{R}(s) = 
\begin{pmatrix}
\mathbf{A} & \mathbf{B}\\
\mathbf{C} & \mathbf{D}
\end{pmatrix}.
\end{equation}
Here $\mathbf{A},\mathbf{B},\mathbf{C}$ and $\mathbf{D}$ are $2\times2$
matrices. In the absence of damping, $\mathbf{R}$ satisfies the
symplecticity condition
\begin{equation}\label{symplectic}
\mathbf{R}^T\mathbf{S}\mathbf{R}=\mathbf{S},
\end{equation}
in which, $\mathbf{S}$ is the
symplectic matrix
\begin{equation}\label{symJ}
\mathbf{S} = \left(
\begin{array}{rrrr} 
0 & 1 & 0 & 0\\
-1 & 0 & 0 & 0\\
0 & 0 & 0 & 1\\
0 & 0 & -1 & 0
\end{array} \right).
\end{equation}
Eq.~\eqref{symplectic} constrains the number of independent elements of
$\mathbf{R}$ to be 10.

In Lie algebra language, $\mathbf{R}$ can be interpreted as a quadratic
Lie generator~\cite{Dragt:1981pj,Chao:2002st} 
\begin{equation}\label{f2}
f_2=\sum_{\begin{array}{c}k+l+m+n=2\\k,l,m,n\ge{}0\end{array}}C_{klmn}x^kp_x^ly^mp_y^n.
\end{equation}
Thus, the transformation of $\vec{v}$ through the matrix $\mathbf{R}$ is
equivalent to a exponential Lie map transformation,
\begin{equation}\label{R_Lie}
\vec{v}^{(n+1)} = R(s)\vec{v}^{(n)}\; \leftrightarrow \; v_i^{(n+1)} =
e^{:f_2:}v_i|_{\vec{v}=\vec{v}^{(n)}}.
\end{equation}
Here, $v_i$ is the $i^{th}$ component of $\vec{v}$. There are also 10
independent quadratic terms in $f_2$, which correspond the 10 independent
elements in $\mathbf{R}$.

The coefficients of monomial terms in $f_2$ are determined by solving a
symmetric, positive definite matrix $\mathbf{F}$ from
\begin{equation}\label{esf=r}
e^{\mathbf{SF}} = \mathbf{R}.
\end{equation}
And the Lie generator $f_2$ reads as
\begin{equation}\label{zfz}
\begin{array}{lll}
f_2 & = & -\frac{1}{2}\vec{v}^T\mathbf{F}\vec{v} \\
& = & f_2^{(0)}+f_2^{(c)} \\
& = & C_{2000}x^2+C_{1100}xp_x+C_{0200}p_x^2\;+ \\
& & C_{0020}y^2+C_{0011}yp_y+C_{0002}p_y^2\;+ \\
& & C_{1010}xy+C_{1001}xp_y+C_{0110}p_xy+C_{0101}p_xp_y.
\end{array}
\end{equation}
Here, $f_2^{(0)}$ is the uncoupled generator. $f_2^{(c)}$ is the linear
coupling generator, which includes four plane-crossing terms,
$xy,\;p_xy,\;xp_y$ and $p_xp_y$. The coefficients of $f_2^{(c)}$ terms
actually characterize the linear coupling between two planes. Our
algorithm is to extract the coefficients of $f_2^{(c)}$ from TbT data,
then to minimize them with non-dispersive skew quadrupoles directly.

Based on previous analyses accomplished by others, some parameterizations
can be derived from the coupled Lie generator $f_2$. Here we briefly
discuss how these parameterizations are related to $f_2^{(c)}$. In
Sect.\ref{exper}, we will compute these parameters using TbT data before
and after correction to demonstrate that our algorithm is equivalent to
these previous analyses.

First, the fully-coupled matrix $\mathbf{R}$ and the Lie generator $f_2$
can be converted through Eq.\eqref{R_Lie} directly. The coupling strength
can be observed from two non-zero off-diagonal blocks $\mathbf{B}$ and
$\mathbf{C}$. Quantitatively Edwards and Teng normalized $\mathbf{R}$ to a
block-diagonal normal mode format~\cite{Edwards:1973fm}. The coupling can
be characterized by a $2\times2$ symplectic matrix $\mathbf{D}$ and a
phase $\phi$.

Mais and Ripken~\cite{Mais:1982hx} proposed another parameterization with
four generalized eigenvectors of $\mathbf{R}$. In this case two modes $I$
and $II$, and four $\beta$-functions can be derived to describe the
frequency and the envelope functions of betatron oscillation. In each
plane, the betatron oscillation is composed of two linear independent
modes
\begin{multline}\label{4beta}
 u =
 \sqrt{J_{u,I}\beta_{u,I}}\cos(\mu_{u,I}+\psi_{u,I})+ \\
 \sqrt{J_{u,II}\beta_{u,II}}\cos(\mu_{u,II}+\psi_{u,II}),
\end{multline}
where $u=x,y$, $\beta_{I,II}$ and $\mu_{I,II}$ are the generalized
Courant-Snyder betatron envelope functions and phase advances for two
modes $I$ and $II$. $J_{I,II}$ and $\psi_{I,II}$ are constants determined
by initial conditions. They will degenerate to the standard Courant-Snyder
parameterization when coupling vanishes.

The fully coupled Lie generator $f_2$ can be separated into two parts as
Eq.~\eqref{zfz}, uncoupled part $f_2^{(0)}$, and coupled part
$f_2^{(c)}$. First, the uncoupled part $f_2^{(0)}$ can parameterized with
Courant-Snyder normalization as
\begin{multline}\label{f20}
f_2^{(0)} = \frac{\mu_x}{2}(\gamma_xx^2+2\alpha_xxp_x+\beta_xp_x^2) +\\
\frac{\mu_y}{2}(\gamma_yy^2+2\alpha_yyp_y+\beta_yp_y^2) 
= \frac{\mu_x}{2}A_x+\frac{\mu_x}{2}A_y.
\end{multline}
Here, $\mu_{x,y}$ is the betatron phase advance per turn,
$\alpha_{x,y},\beta_{x,y}$ and $\gamma_{x,y}$ are the Twiss parameters at
$s$, and $A_{x,y}$ are the action variables. Then four coupled terms,
\begin{equation}\label{f2c_1}
f_2^{(c)} = C_{1010}xy+C_{1001}xp_y+C_{0110}p_xy+C_{0101}p_xp_y
\end{equation}
can be expressed as a summation of the resonance basis of
$f_2^{(0)}$
\begin{equation}\label{f2c_2}
f_2^{(c)} = \sum_{a+b=1,c+d=1}h_{abcd}|abcd\rangle.
\end{equation}
Where
\begin{multline}\label{resonancebasis}
|abcd\rangle = \\
(\sqrt{A_x}e^{i\phi_x})^a(\sqrt{A_x}e^{-i\phi_x})^b
(\sqrt{A_y}e^{i\phi_y})^c(\sqrt{A_y}e^{-i\phi_y})^d
\end{multline}
are the resonance basis of non-coupled $f_2^{(0)}$ in
Eq.~\eqref{f20}. $A_{x,y}$ and $\phi_{x,y}$ in Eq.~\eqref{f20} and
\eqref{resonancebasis} are the action-angle canonical variables. Thus, two
pairs of complex conjugates coefficients characterize the coupling
\begin{equation}\label{h1010h1001}
\begin{matrix}
h_{1010} & = & h_{0101}^*, \\
h_{1001} & = & h_{0110}^*,
\end{matrix}
\end{equation}
in which $h_{1010}$ is referred as linear sum resonance driving
term~(RDT), because it can drive a sum resonance when the system tune is
close to the resonance line $\nu_x+\nu_y=n$. $h_{1001}$ is defined as
difference RDT for the same reason. Ref.~\cite{Calaga:2005bn} proves that
the RDTs can be merged with Edwards-Teng parameterization
~\cite{Edwards:1973fm}.

Assuming the system is decoupled at a certain observation point $s$, in
the matrix language, two off-diagonal blocks $\mathbf{B}=\mathbf{C}=0$ in
$\mathbf{R}$; in Lie algebra language, four plane-crossing terms disappear
from $f_2$ of Eq.~\eqref{f2}; and $\beta_{x,II}$ and $\beta_{y,I}$ in
Mais-Ripken's parameterization also degenerate to zeros, so are RDTs
in~\eqref{h1010h1001}. In other words, the off-diagonal matrices
$\mathbf{B}$ and $\mathbf{C}$ in $\mathbf{R}$, the crossing terms in
$f_2$, two RDTs, and two coupling $\beta$-functions characterize a common
physics quantity - linear coupling observed at the location of this
specific position $s$. The goal of linear coupling control is to minimize
these non-zero terms.

It is worthwhile to point out that, even though a system is decoupled at
one observation point, it is not necessarily decoupled at another
one. This is clear since $f_2^{(c)}$ is $s$-dependent. For ring-based
light sources, it is crucial to control the coupling at insertion device
locations.

\section{\label{measurement}Experimental characterization of linear coupling}
In this section, we discuss how to characterize the coupling with BPMs'
TbT readings experimentally, some other techniques can be found
in~\cite{Sagan_Rubin_1999,Huang_Sebek_Martin_2010}. In order to obtain
synchronized and consecutive TbT data, the beam needs to be excited by
pulse magnets, then all BPM readings must be timed with the pulse magnets
triggering event within one revolutionary period. From the TbT data array,
we first choose two neighboring BPMs, $P_i$ and $P_{i+1}$, with only a few
magnets in-between, and assume the linear transforming matrix
$\mathbf{M}_{i,i+1}$ between these two BPMs is known. By ignoring damping
and de-coherence, the two BPMs' readings (after subtracting the closed
orbit) at the $n^{th}$ turn are related by $\mathbf{M}_{i,i+1}$
\begin{equation}\label{Mii+1}
\begin{pmatrix}
x^{(n)} \\
p_x^{(n)}\\
y^{(n)} \\
p_y^{(n)}
\end{pmatrix}_{i+1}
= \mathbf{M}_{i,i+1} 
\begin{pmatrix}
x^{(n)} \\
p_x^{(n)}\\
y^{(n)} \\
p_y^{(n)}
\end{pmatrix}_i.
\end{equation}
With Eq.~\eqref{Mii+1}, $p_x^{(n)},p_y^{(n)}$ at both BPMs are
determined. Therefore we obtain beam coordinates in phase space at the
locations of the two BPMs for multiple turns. Then the one-turn-map at the
location of the $i^{th}$ BPM is the least-squares solution of the linear
equations for multiple turns.
\begin{equation}\label{lstsqu}
\begin{pmatrix}
x^{(n)} & \hdots & x^{(2)} \\
p_x^{(n)} & \hdots & p_x^{(2)}\\
y^{(n)} & \hdots & y^{(2)} \\
p_y^{(n)} & \hdots & p_y^{(2)}
\end{pmatrix}_{i}
= \mathbf{R}_{i} 
\begin{pmatrix}
x^{(n-1)} & \hdots & x^{(1)} \\
p_x^{(n-1)} & \hdots & p_x^{(1)}\\
y^{(n-1)} & \hdots & y^{(1)} \\
p_y^{(n-1)} & \hdots & p_y^{(1)}
\end{pmatrix}_i.
\end{equation}

\begin{figure}[!ht]
\centering
\includegraphics[width=0.8\columnwidth]{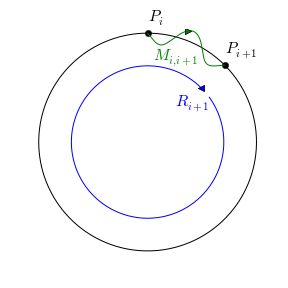}
\caption{\label{twoBpm}Using two neighboring BPMs reading and the
  transport matrix $M_{i,i+1}$ to construct beam coordinates in phase
  space, then fit out the one-turn-map $\mathbf{R}$ at each observation
  BPM.}
\end{figure}

In principle, two consecutive turns data can uniquely define a one-turn
map. But due to various errors from BPMs readings, magnet power suppliers
jittering, and etc., we have to fit multiple consecutive turns data with
the least square method to filter those random errors out. Usually more
than 500 turns data are used to solve Eq.~\eqref{lstsqu}.

As mentioned before, the couplings seen at different BPMs locations could
be different. For a ring-based light source unless for special purpose,
e.g. increasing beam volume for longer Touschek life time, it is
preferable to have no coupling all over the storage ring, especially at
the source points where insertion devices are located. Thus we need to fit
out the one-turn-maps at multiple BPM pairs by applying Eq.~\eqref{Mii+1}
and \eqref{lstsqu} repeatedly.

\begin{figure}[!ht]
\centering
\includegraphics[width=0.8\columnwidth]{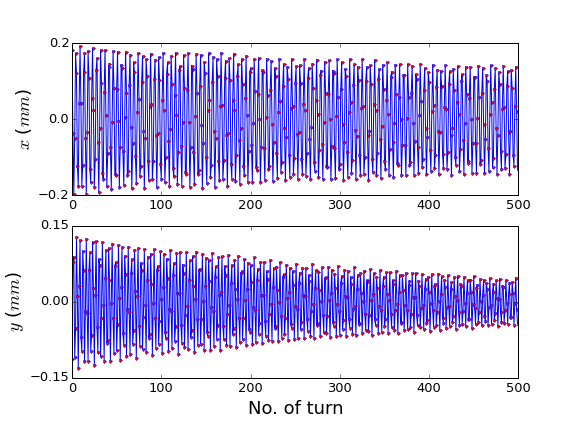}
\caption{\label{TbT}A set of typical TbT data observed by a BPM on the
  NSLS-II ring. A strong amplitude variation, especially in the vertical
  plane, is visible. With damping wigglers gaps closed the damping rate
  will be further enhanced.}
\end{figure}

There existes another method to extract the N-turn map to avoid computing
$p_{x,y}$ in hadron rings ~\cite{Fischer_2003}. In the NSLS-II storage
ring, a strong amplitude variation due to radiation damping, or
decoherence is visible (see Fig.~\ref{TbT}) from TbT data. In order to
mitigate this effect, we choose two neighboring BPMs to reconstruct
momenta $p_{x,y}$, then two consecutive turns data to extract one-turn
maps.

Once $\mathbf{R}$ is obtained, the coefficients of coupling terms
$C_{klmn}$, and then RDTs $h_{abcd}$ can be calculated with
Eqs.~\eqref{esf=r},~\eqref{zfz} and ~\eqref{f2c_2} respectively. The two
coupling $\beta$-functions can be calculated with the approach explained
in \cite{Willeke:1988zu}.

One thing needs to be emphasized here is, the direct measured $\mathbf{R}$
with Eq.~\eqref{lstsqu} is not always exactly symplectic due to various
measurement errors. A symplectic matrix $\mathbf{R_s}$ can be obtained in
the following way. First, $\mathbf{R}$ can be approximated to a Lie
generator $f_2$ using Eq.~\eqref{esf=r}. Then we can act $f_2$ on each
canonical variable to get one row of a symplectic matrix $\mathbf{R_s}$ as
explained in ref.~\cite{Chao:2002st},
\begin{multline}\label{equ:symplectification}
x_1 = e^{:f_2:}x|_{x=x_0,y=y_0,p_x=p_{x0},p_y=p_{y0}} \\ 
= R_{s,11}x_0+R_{s,12}p_{x0}+R_{s,13}y_0+R_{s,14}p_{y0}.
\end{multline}
Here, only the first row is listed, other three rows can be obtained in
the same way. An alternative way of symplectifying $\mathbf{R}$ is given in
ref. ~\cite{Fischer_2003}.

\begin{figure}[!ht]
\centering
\includegraphics[width=\columnwidth]{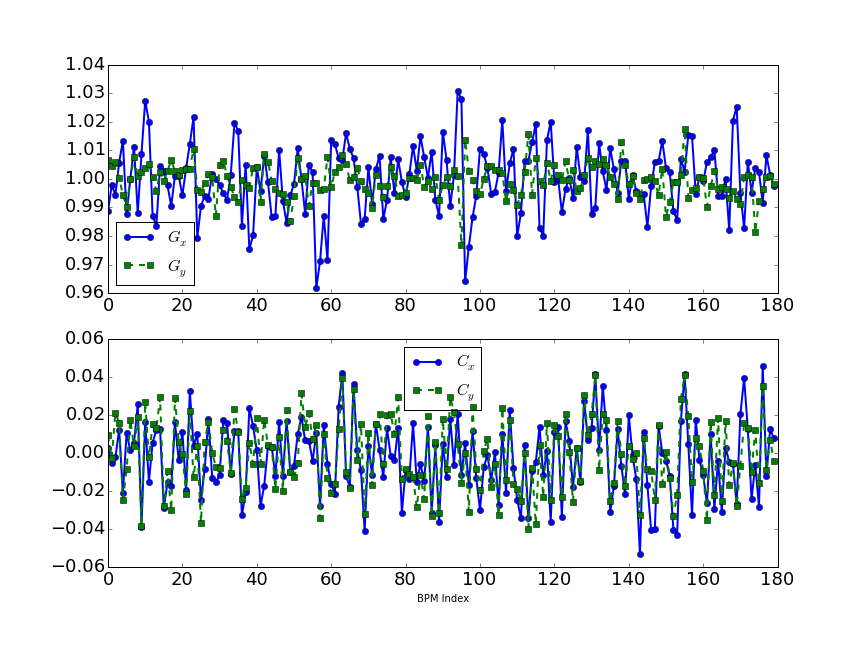}
\caption{\label{fig:bpmCali}Gain (upper) and coupling (lower) calibration
  coefficients for 180 BPMs}
\end{figure}

Now we discuss the control of various measurement errors. First, BPM's
imperfections can affect the calculation of $f_2^{(c)}$ and therefore
$C_{klmn}$. In order to mitigate these affects, for each BPM, four
parameters fitted by the LOCO \cite{Safranek_1997, Safranek_2007,
  Yang_2007, xyang} method give the full linear transformation between the
raw TbT readings $(\bar{x},\bar{y})$ and the realistic beam trajectory
$(x,y)$:
\begin{equation}\label{equ:bpm_cali}
\begin{pmatrix}
\bar{x} \\
\bar{y}
\end{pmatrix}
=
\begin{pmatrix}
G_x & C_x\\
C_y & G_y
\end{pmatrix}
\begin{pmatrix}
x \\
y
\end{pmatrix},
\end{equation}
where, $G_{x,y}$ are the gain calibrations, and $C_{x,y}$ are the coupling
calibrations due to the roll and the associated construction errors. The
four parameters vary for each BPM, as shown in
Fig.~\ref{fig:bpmCali}. Calibrated data are obtained by implementing the
inverse transformation of Eq.~\eqref{equ:bpm_cali} on raw data.

BPM resolution is measured as $1\mu{}m$ at 10mA stored beam
current~\cite{Cheng}. During the coupling characterization, we usually
excite beam with an amplitude less than $\pm0.5mm$. This resolution can
satisfy the requirement of precise characterization of linear coupling.

A systematic error comes from the assumption that the transforming matrix
$\mathbf{M}_{i,i+1}$ between two BPMs is known in Eq.~\eqref{Mii+1}. In
order to minimize the effect caused by unknown magnets errors, we
intentionally choose two BPMs separated by as few as possible magnets. In
the NSLS-II storage ring, only those BPM pairs separated by a long drift
($l>5.5m$) and two sextupoles are used. The closed orbit has been aligned
to the center of relevant quadrupoles
~\cite{Portmann_Robin_Schachinger_1995} and the closed orbit displacements
at the locations of sextupoles are negligible, thus $\mathbf{M}_{i,i+1}$
is simplified as a coordinate transformation through a drift. The presence
of sextupoles intrinsically affects $p_x$ and $p_y$ as a systematic
error. A detailed simulation was set up to study this effect. A set of
simulated TbT data was obtained by tracking a particle with a similar
transverse amplitude ($\pm0.5mm$) as we used in the experiment. Then we
use momenta $p_{x,y}$ in the simulated data to compute the one-turn map
$\mathbf{R}_{0}$ directly, and compare it against the map $\mathbf{R}_{1}$
from the momenta reconstructed by the coordinates $x,y$ with the
approximated linear transfer map. Two obtained maps are shown as,
\begin{equation}\label{R_sim}
\mathbf{R}_{0} = \\ \left( \begin{array}{rr|rr}
-1.7009  &  8.6528 & -0.2410 &  0.8301 \\  
-0.5311  &  2.1140 & -0.0617 &  0.2120 \\ \hline
-0.0305  & -0.0446 & -3.1178 & 12.1421 \\ 
-0.0009  & -0.0043 & -0.8506 &  2.9920 \\
\end{array} \right),
\end{equation}
and
\begin{equation}\label{R_exp}
\mathbf{R}_{1} = \\ \left( \begin{array}{rr|rr}
-1.7005 &  8.6492 & -0.2405 &  0.8287 \\  
-0.5310 &  2.1135 & -0.0615 &  0.2117 \\ \hline 
-0.0305 & -0.0448 & -3.1180 & 12.1419 \\
-0.0009 & -0.0044 & -0.8507 &  2.9922 \\
\end{array} \right).
\end{equation}
Consider betatron oscillation amplitude is around $\pm0.5mm$, then the
difference of TbT readings between these two maps is below
$~0.5\mu{}m$. The NSLS-II BPM resolution is around $1\mu{}m$. Therefore,
in this particular case, the sextuple effect on TbT data is too weak to be
measured experimentally.

In electron storage rings, a systematic error is due to radiation and
collective damping. Radiation damping is negligible because we fit the
data between two consecutive turns. Collective damping can be mitigated by
filling the ring with a low charge per bunch, while BPMs still have reliable
readings.

Another error is the decoherence effects caused by non-zero chromaticity
and nonlinearity~\cite{Meller:1987ug}. They can be effectively suppressed
by adjusting chromaticities close to zeros, and using low beam
excitations separately.

As usual, multiple TbT data collection are repeated for each scenario to
suppress any random uncertainty and also are used to estimate measurement
error fluctuation. The achieved fluctuations of $C_{klmn}$ for multiple
measurements are around 5\% (see ~Fig.\ref{Ccomp}).

\section{\label{algorithm}Correction algorithm}

Based on the designed lattice model, four plane-crossing terms' dependence
on non-dispersive skew quadrupoles are calculated with
\begin{equation}\label{eq:respmat}
N_{klmn,i,j}=\frac{\partial C_{klmn,i}}{\partial K_j},
\end{equation}
where $C_{klmn,i}$ are the coefficients of Eq.~\eqref{f2c_1} observed at
the location of $i^{th}$ BPMs, $K_j$ is the $j^{th}$ skew quadrupole
normalized focusing strength, $N_{klmn,i,j}$ is the $i^{th}$ BPM's
dependence on the $j^{th}$ skew quadrupole. Once these four coefficients
in Eq.~\eqref{f2c_1} at each observation location are measured, the needed
skew quadrupole corrections to minimize them are obtained by iteratively
solving the following linear equations
\begin{equation}\label{dk_solve}
\Delta\vec{\mathbf{C}}_{klmn} = \mathbf{N}_{klmn} \Delta\vec{\mathbf{K}}.
\end{equation}
Since there are four goal functions per observation location, we
vertically stack their response matrices $\mathbf{N}_{klmn}$ with
different weights to minimize them simultaneously.

\section{\label{exper}Experiment on NSLS-II storage ring}

In order to control the vertical beam size, inside each cell of the
NSLS-II storage ring, a skew quadrupole is incorporated into the
lattice. In the odd numbered cells, it is located in the dispersive
region, which can be used for vertical dispersion compensation. And in the
even numbered cells, it is located in non-dispersive regions, which can be
used for linear coupling correction. With this configuration, we can
correct the vertical dispersion first, and then the linear
coupling. Actually we can combine these two corrections with a weighted
and combined response matrix to correct them simultaneously as suggested
in Ref.~\cite{Franchi_2011}. It will become necessary if a different
lattice configuration with non-zero dispersion all around the ring. 

\begin{figure}[!ht]
\centering
\includegraphics[width=\columnwidth]{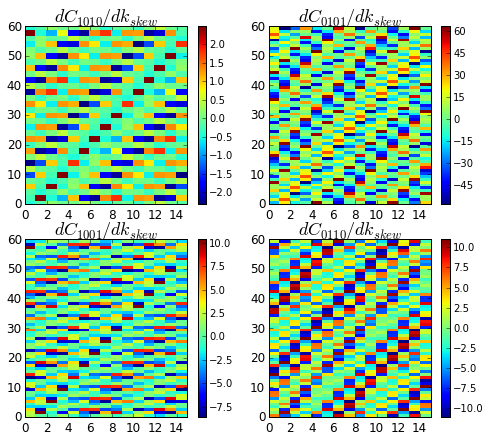}
\caption{\label{fig:respmat}Response matrices of the coupling coefficients
  in Eq.~\eqref{f2c_1} dependence on non-dispersive skew quadrupole
  strength. The horizontal axes are skew quadrupole's index, and the
  vertical axes are BPM's index.}
\end{figure}

In this paper, we only focus on the coupling correction. A total of 60
BPMs located in 30 straight sections are used to characterize the
coupling. Therefore, each individual response matrix is $60\times15$, as
illustrated in Fig.~\ref{fig:respmat}. The dimension of stacked response
matrix is $240\times15$.

Eq.~\eqref{dk_solve} is solved with SVD algorithm. At each iteration,
we only keep those singular values greater than 50\% of the largest one in
order to minimize corrector strengths.

\begin{figure}[!ht]
\centering \includegraphics[width=\columnwidth]{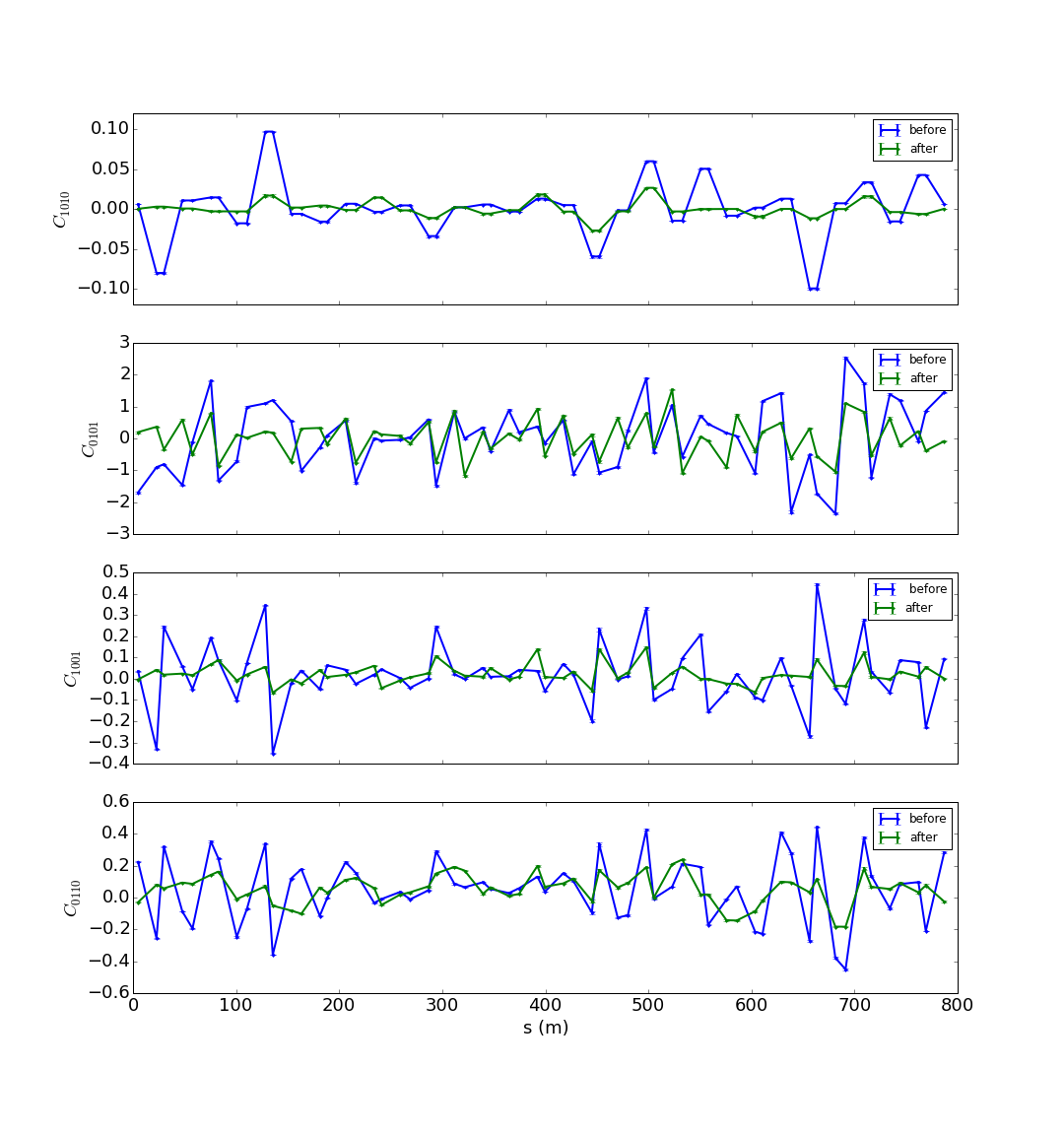}
\caption{\label{Ccomp}Comparison of the four linear coupling coefficients
  before and after 3 iterative corrections. The error bars are the
  standard deviations of 10 turn-by-turn data snapshots.}
\end{figure}

Fig.~\ref{Ccomp} shows the comparison of four coupling coefficients
$C_{1010}$, $C_{1001}$, $C_{1001}$ and $C_{0110}$ before and after
applying correction. Among them, the suppression on $C_{1010}$ and
$C_{1001}$ are obvious at most BPMs.

As explained in Sect.~\ref{coupling}, several other parameterizations are
used to characterize the coupling. Next we calculate and compare their
parameters before and after applying correction. It will confirm that, the
minimization of the crossing terms in the Lie generator is equivalent to
other approaches.

\subsection{Linear one-turn-map}
The measured one-turn-map seen by one of BPMs before applying the
correction reads as
\begin{equation}\label{R_0}
\mathbf{R}_0 = \left( \begin{array}{rr|rr}
-0.1005 & 22.5204 & \mathbf{-0.1943} & \mathbf{1.5475} \\
-0.0471 & 0.5404 & \mathbf{-0.0030} & \mathbf{0.0590} \\ \hline
\mathbf{0.0148} & \mathbf{1.2383} & -1.6854 & 11.0648 \\
\mathbf{0.0079} & \mathbf{0.1962} & -0.2976 & 1.3562
\end{array} \right).
\end{equation}
And after three iterative corrections, it becomes
\begin{equation}\label{R_1}
\mathbf{R}_1 = \left( \begin{array}{rr|rr}
-0.1029 & 22.5040 & \mathbf{0.0197} & \mathbf{-0.0673} \\
-0.0469 & 0.5394 & \mathbf{-0.0004} & \mathbf{0.0020} \\ \hline
\mathbf{0.0050} & \mathbf{-0.0780} & -1.6765 & 11.0022 \\
\mathbf{0.0003} & \mathbf{-0.0022} & -0.2958 & 1.3447 
\end{array} \right).
\end{equation}
The off-diagonal elements after correction in Eq.~\eqref{R_1} are found to
be significantly smaller than before correction in Eq.~\eqref{R_0}. It
means that the coupling matrix $\mathbf{D}$ and $\phi$ in Edwards-Teng
parameterization has been reduced successfully.

\subsection{Coupling $\beta$-functions}
\begin{figure}[!ht]
\centering
\includegraphics[width=\columnwidth]{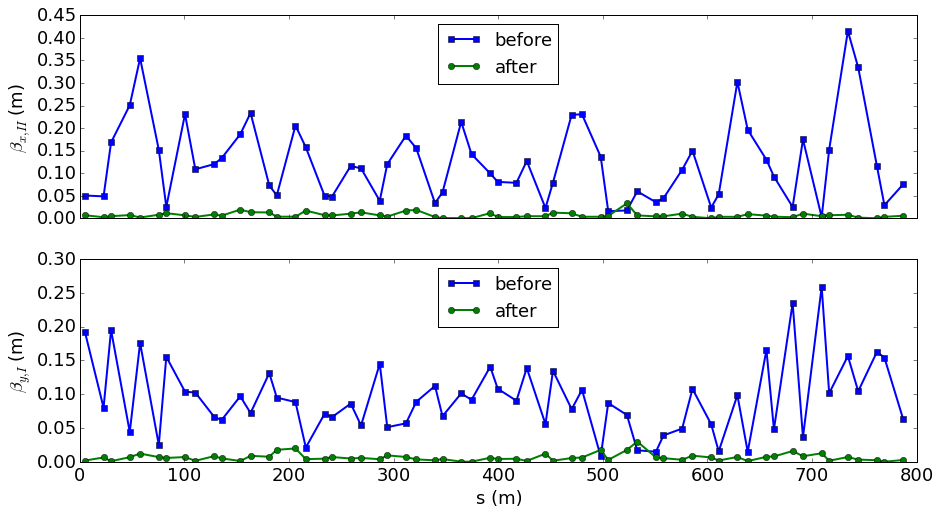}
\caption{\label{coupledbeta}Comparison of the coupling $\beta$-functions
before and after 3 iterative corrections.}
\end{figure}

In Mais-Ripken's parameterization, besides two dominant $\beta_{x,I}$ and
$\beta_{y,II}$, another two small envelope functions $\beta_{x,II}$ and
$\beta_{y,I}$ actually represent the coupling motion between two
planes. They were calculated from the measured $\mathbf{R}$. The
comparison is illustrated in Fig.~\ref{coupledbeta}. Both the maximum and
average values of the two coupling $\beta$-functions are suppressed
significantly after three iterations.

\subsection{Resonance driving terms}
\begin{figure}[!ht]
\centering
\includegraphics[width=\columnwidth]{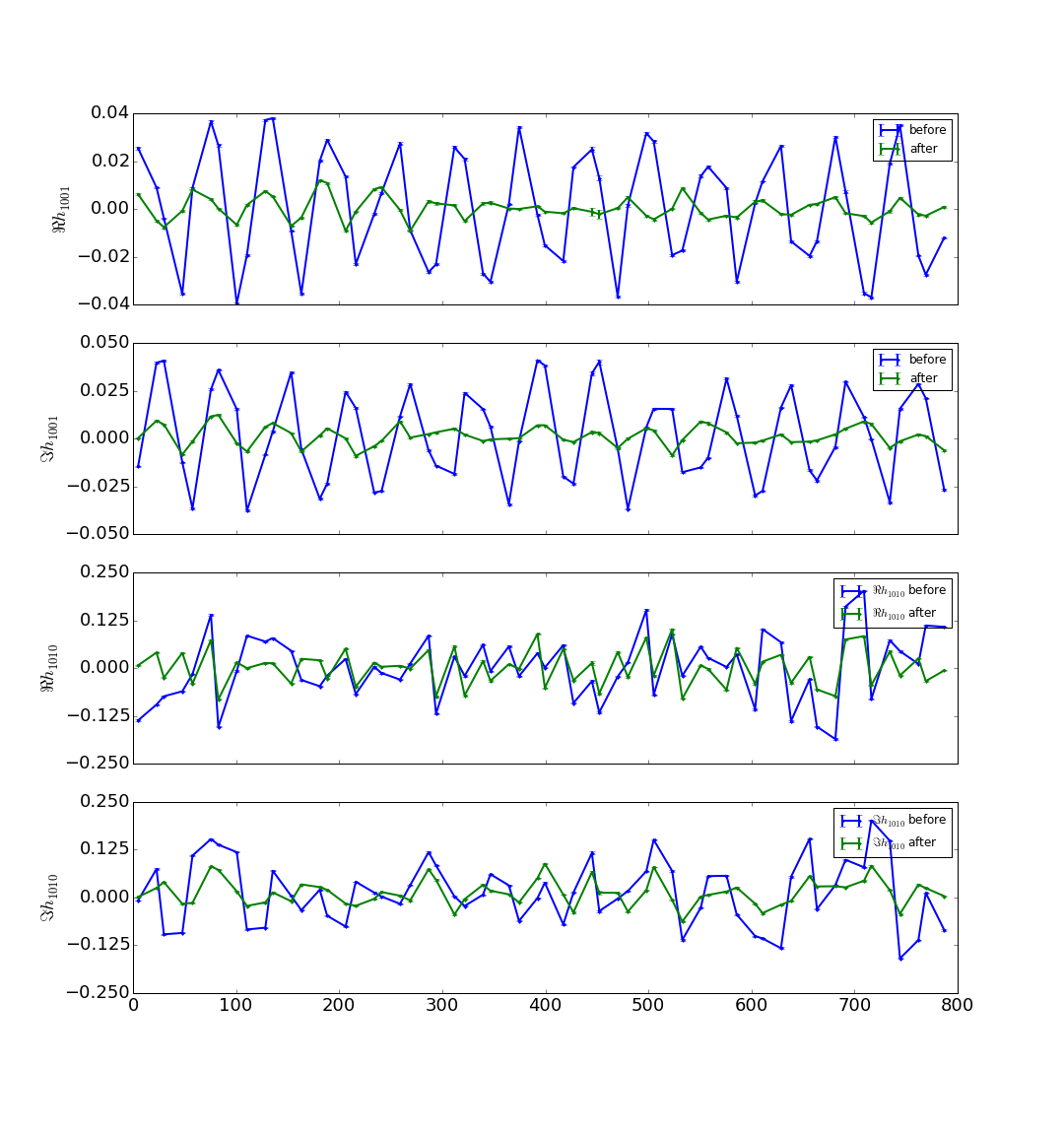}
\caption{\label{hcomp}Comparison of the RDTs $h_{1001}$ and $h_{1010}$before
  and after 3 iterative corrections. The upper two plots are the real and
  imagery parts of the difference resonance, and the lower two plots are
  for the sum resonance.}
\end{figure}

Since the tune of NSLS-II ring tune is close to the difference resonance
line $\nu_x-\nu_y=17$, the suppression on RDT $h_{1001}=h_{0110}^*$ are
relatively dramatic. Comparison of its real and imagery parts before and
after corrections are illustrated as the $1^{st}$ and $2^{nd}$ plot in
Fig.~\ref{hcomp} respectively. In the meantime, the suppression on the sum
RDT $h_{1010}=h_{0101}^*$ is also visible as shown in the $3^{rd}$ and
$4^{th}$ plots of Fig.~\ref{hcomp}. It is also possible to correct the sum
and difference together using a weighted response matrix of RDTs versus
skew quadrupoles ~\cite{Franchi_2011}.

\subsection{Spectra of Betatron oscillation}
\begin{figure}[!ht]
\centering
\includegraphics[width=\columnwidth]{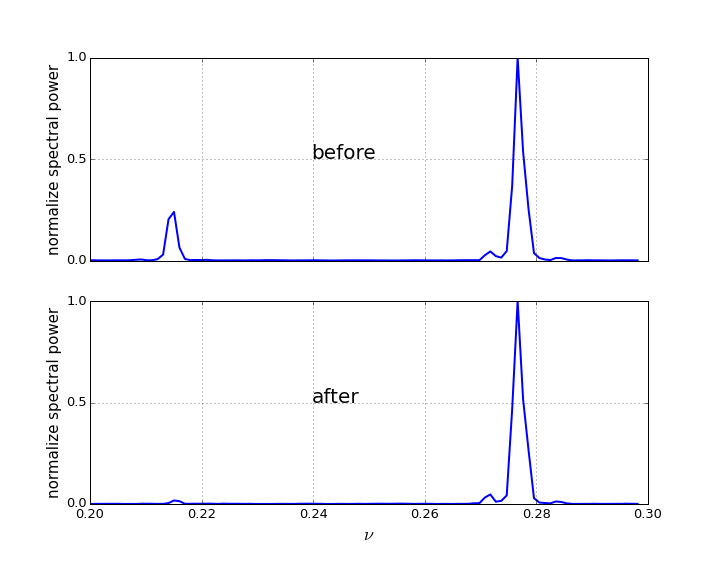}
\caption{\label{spectrum}The vertical TbT spectra before (upper) and after
  (lower) coupling correction. After correction, the horizontal mode
  frequency peak is suppressed in the vertical plane.}
\end{figure}

The vertical spectra of the TbT data before and after coupling correction
are illustrated in Fig.~\ref{spectrum}. Before correction, a horizontal
betatron mode frequency peak is clearly visible in the vertical
spectrum. After applying correction on skew quadrupoles, it has been
effectively suppressed.

\subsection{Beam transverse profile}
\begin{figure}[!ht]
\centering
\includegraphics[width=\columnwidth]{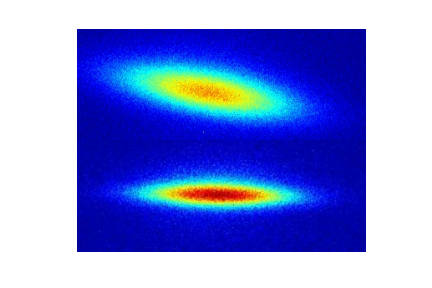}
\caption{\label{beamprofile}Comparison of beam profiles taken by the
  pinhole camera. The upper image was taken before correction, and the
  lower one was after correction. The effective vertical emittance was
  suppressed from 32.0$pm\cdot{}rad$ to 6.4$pm\cdot{}rad$.}
\end{figure}

Another direct observation of the linear coupling is beam transverse
profile - its tilt angle and vertical size. A x-ray diagnostic beam line
configured with a pinhole camera provides the real time beam profiles and
its 2D Gaussian fitting dimensions. The CCD camera pixel dimension with
respect to beam size image has been well calibrated as $0.449\mu{}m/pixel$
by displacing beam at different orbits. Before the coupling correction,
beam transverse profile has a tilted angle with the 32$pm\cdot{}rad$
vertical beam emittance. After 3 times iterations, Both the tilt angle and
vertical beam size become much smaller than before correction. The profile
comparison before and after correction observed by the pinhole camera is
shown in Fig.~\ref{beamprofile}.

In the meantime, the phenomenon of beam vertical size reduction was
observed at another source point, the Hard X-ray Nano-probe beam line
in-vacuum undulator ~\cite{Oleg}. As we mentioned before, our strategy is
to minimize the coupling at multiple observation points, especially the
locations where insertion devices are located.

\subsection{Beam lifetime}
\begin{figure}[!ht]
\centering
\includegraphics[width=\columnwidth]{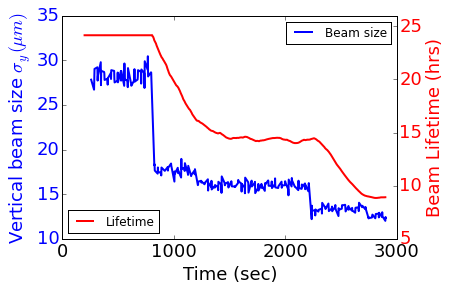}
\caption{\label{lifetime}Beam lifetime changes with vertical beam size
  during the vertical dispersion and the coupling correction. The first
  lifetime drop-off is due to the dispersion correction, and the
  subsequent drop-offs are caused by the iterative coupling
  corrections. The storage ring was filled with 5.5mA in 200
  buckets. Lifetime was poor because the vacuum had not been well
  conditioned with high beam current during the early stage of
  commissioning.}
\end{figure}

The Touschek lifetime depends on the vertical beam size linearly. The
existence of both vertical dispersion and linear coupling can blow up beam
size vertically. Therefore, as the corrections were implemented, the
beam lifetime was observed to be linearly scaled with the vertical beam
size as illustrated in Fig.~\ref{lifetime}.

\section{\label{summary}Summary}
From beam turn-by-turn BPM readings, four plane-crossing terms in Lie
generators are extracted to characterize linear coupling directly. And
coupling control can be realized by utilizing their dependence on the
non-dispersive skew quadrupoles. In the meantime, several other
parameters, such as, Edwards-Teng symplectic matrix normalization,
resonance-driving terms, and Mais-Ripken $\beta$-functions can be derived
from the Lie generators. The application of this approach on the NSLS-II
ring successfully control the coupling and reduce the vertical emittance
below the diffraction limit. The effectiveness of our coupling correction
method has been verified by using another method - independent component
analysis~(ICA), and reported in~\cite{Huang_2015}.

Some existing tools, such as LOCO ~\cite{Safranek_1997, Safranek_2007} can
realize linear coupling correction as a small part of the overall
framework. But to measure an orbit response matrix, and then to fit it to
a lattice model take some time, which usually needs a dedicated beam study
time. A major benefit of using TbT data to characterize and control linear
coupling is that it can be automatized as a feedback process during
routine operation, because it can accomplish data access and correction
within several seconds.

\begin{acknowledgments}
Work was supported by the U.S. Department of Energy, Office of Science,
Office of Basic Energy Sciences, under Contract No. DE-AC02-98CH10886.
\end{acknowledgments}

\bibliography{tbt_coupling_ref}
\end{document}